\documentclass{article}
\usepackage{spconf,amsmath,graphicx}
\usepackage{tabularx}
\usepackage{multirow}
\usepackage{hyperref}
\usepackage[binary-units]{siunitx}

\usepackage{adjustbox}
\usepackage{array}
\usepackage{subfigure}
\usepackage{fixltx2e}
\usepackage{fix2col}
\newcolumntype{R}[2]{%
    >{\adjustbox{angle=#1,lap=\width-(#2)}\bgroup}%
    l%
    <{\egroup}%
}
\newcommand*\rot{\multicolumn{1}{R{30}{1em}}}

\title{WaveGlove: Transformer-based hand gesture recognition using multiple inertial sensors}
\name{Matej Králik, Marek Šuppa}
\address{Faculty of Mathematics, Physics and Informatics, Comenius University Bratislava, Slovakia}
\begin{document}
%
\maketitle
\begin{abstract}

Hand Gesture Recognition (HGR) based
on inertial data has grown considerably in recent years, with the state-of-the-art approaches utilizing
a single handheld sensor and a vocabulary comprised of simple gestures.
\
In this work we explore the benefits of using multiple inertial sensors. Using \textsc{WaveGlove}, a custom hardware prototype 
in the form of a glove with five inertial sensors, we acquire two datasets consisting of over $11000$ samples.
\
To make them comparable with prior work, they are normalized along with $9$ other publicly available datasets, and subsequently used to evaluate a range of Machine Learning approaches for gesture recognition, including a newly proposed Transformer-based architecture.
Our results show that even complex gestures involving different fingers can be recognized with high accuracy.
\
An ablation study performed on the acquired datasets demonstrates the importance of multiple sensors, with an increase in performance when using up to three sensors and no significant improvements beyond that.


\end{abstract}

\begin{keywords}
hand gesture recognition, transformer, multi-sensor, inertial sensor
\end{keywords}
\section{Introduction}
\label{sec:intro}

Human-computer interaction is experiencing a significant increase in
popularity due to the high availability and low cost of various sensors.
Human Activity Recognition (HAR) is most often performed using
vision-based sensors, which often need to be pre-installed
in the environment, require favorable light conditions,
specific position and rotation of the user and significant computing power.
Inertial Measurement Units (IMUs) are a type of inertial sensor that provides a more pervasive experience.
IMUs have lower power consumption, tend to be smaller and can work in a 
variety of environments.

Hand Gesture Recognition (HGR) -- a sub-field of HAR -- is of particular interest,
because hand gestures are a natural method of expression for humans.
Practical applications of HGR include
device control~\cite{12_he_jin_cellphones},
health monitoring, gaming, virtual and augmented reality
and biometrics~\cite{gait-biometry}.
The state-of-the-art approaches utilize a single (often handheld) IMU
sensor and a vocabulary of very simple gestures
\cite{27_gesture_keeper,30_online_rnn,06_costante_dataset,05_uwave}.
The use of multiple sensors is largely unexplored,
which can be attributed to their significant cost in the past.
At the same time,
various hand gestures used on a daily basis make distinctive use of each of the fingers,
making it hard to distinguish
these natural gestures using only one handheld sensor.

\begin{figure}[t]
  \centering
  \includegraphics[width=\linewidth]{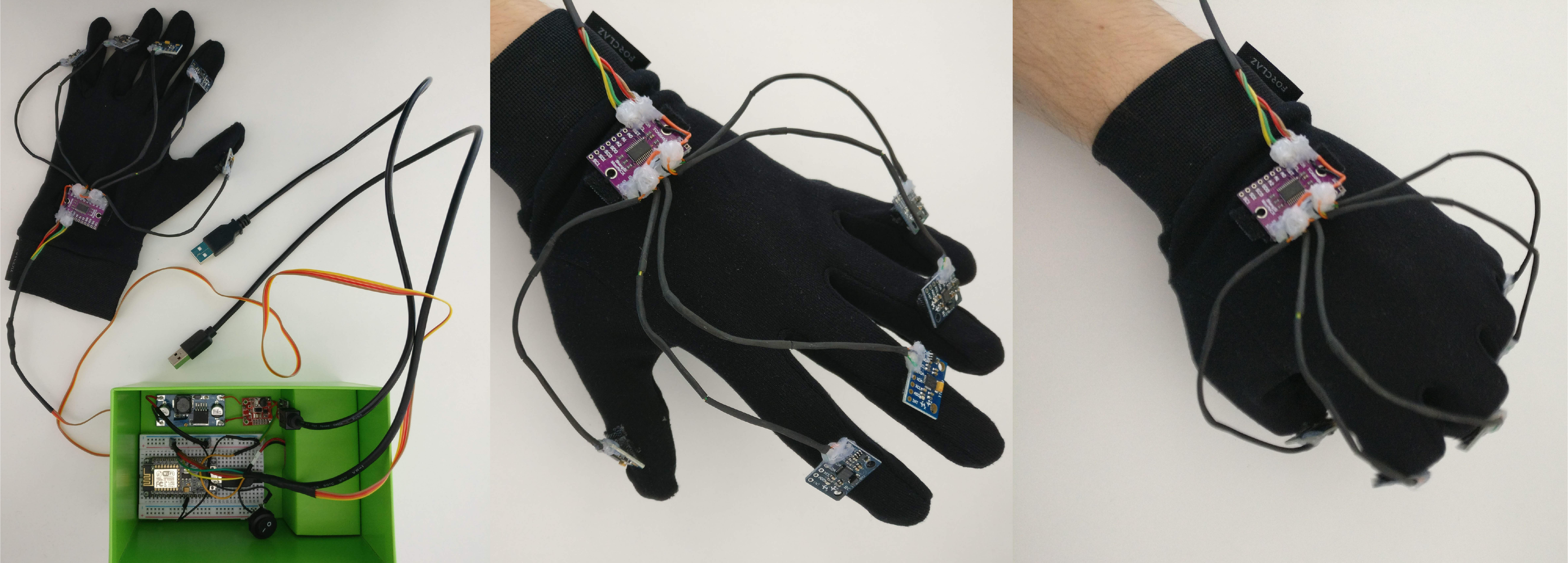}
  \caption{The \textsc{WaveGlove} prototype.}
  \label{fig:waveglove}
\end{figure}

To explore the multi-sensor scenario we built a custom hardware prototype
called \textsc{WaveGlove} and designed two gesture vocabularies:
\textsc{WaveGlove}-single with $8$ whole-hand movements and \textsc{WaveGlove}-multi with  $10$ carefully designed gestures, which exhibit different movements for each of the fingers.
Using the prototype we acquired $1000$ and $10000$ samples for datasets based on the
\textsc{WaveGlove}-single and \textsc{WaveGlove}-multi vocabularies, respectively.

In an effort to put our work in context, we use a set of $11$ datasets.
It includes $2$ datasets captured with \textsc{WaveGlove},
$3$ publicly available datasets we normalized as part of this work and
$6$ datasets normalized in \cite{41_standardization_sota}. These are used to evaluate a wide selection of Machine Learning approaches for hand gesture recognition. We reproduce multiple
previously published methods and propose an improved approach
based on the Transformer~\cite{attention-is-all},
which achieves state-of-the-art results.

To the best of our knowledge, our work is the first to specifically examine
the impact of using multiple sensors for HGR.
Through an ablation study we explore the accuracy implications of multiple sensors. We also compare the recognition performance of individual sensors
and demonstrate a dependency between sensor position and gesture type.

In summary, we built a custom multi-sensor hardware prototype and
used it to acquire datasets of over $11000$ samples.
We further present $11$ datasets in a uniform, normalized format,
use them to evaluate a number of recognition approaches.
Finally, we propose a Transformer-based model and
report state-of-the-art results in its evaluation on the presented datasets.
Using this model we conduct an ablation study which shows that
increasing the number of sensors leads to increased recognition accuracy, up to a point.


\section{Related work}
\label{sec:related}

Several recent comprehensive surveys summarize research and
define the state of the art in the field of HAR~
\cite{45_sota_challenges,35_overview_challenges}.
The most commonly used recognition methods include Classic Machine Learning,
although Deep Learning methods experience a significant
rise in popularity lately~\cite{42_comprehensive_survey}.

Catal et al.~\cite{method-catal} used classifier ensembles in combination
with hand-crafted features. Convolutional Neural Networks were
applied to the task of HAR in~\cite{method-chenxue} and
\cite{method-hachoi} demonstrated that 2D convolutions 
achieve better results than one dimensional convolutions.
In \cite{30_online_rnn}, recurrent LSTM layers were used for online
recognition and \cite{13_deep_fisher} has achieved state-of-the-art results
using bidirectional LSTM and GRU layers.
DeepConvLSTM~\cite{deepconvlstm}
represents a combination of convolutional and recurrent layers.
Inspired by the Transformer~\cite{attention-is-all}
(which established a new state-of-the-art
for machine translation tasks),
\cite{32_self_attention} introduces the self-attention mechanism in the field of HAR.
Our work builds on these results and proposes an improved method
based on the self-attention mechanism.

Hand Gesture Recognition was previously explored in
\cite{12_he_jin_cellphones, 30_online_rnn, 06_costante_dataset, 05_uwave}.
In \cite{sign-language} a similar glove prototype was built to recognize
french sign language characters by estimating the rotation of each finger.
A single-sensor glove prototype called GestGlove~\cite{40_gestglove} was built
to recognize a simple set of hand gestures allowing phone control.
In \cite{icassp-last-year-forked}, the concept of forked recurrent neural networks
is applied to HGR.

One of the current issues in the field of HAR is the lack of a standardized benchmark. A base for a standardized benchmark with a predefined
methodology and publicly released code has been laid in \cite{41_standardization_sota}.
We build on top of this work and aim to extend the realm of reproducible HAR methods.
From the dataset perspective,
Opportunity~\cite{ds-opportunity1}, Skoda~\cite{ds-skoda1} and PAMAP2~\cite{ds-pamap2}
are three widely used HAR datasets, which use multiple sensors.
In \cite{41_standardization_sota}, $6$ more HAR datasets are identified and normalized.
The uWave~\cite{05_uwave} dataset presents one of the few available datasets which
focus on hand gestures.



\section{WaveGlove}
\label{sec:waveglove}

We built a custom hardware glove prototype called \textsc{WaveGlove}
(shown in Figure \ref{fig:waveglove}).
It is based on a left-hand silken glove to which 5 IMUs
are attached: one MPU6050~\cite{datasheet-mpu6050} module on each of the fingers. A multiplexer is present on the back of the palm, connected via
a flexible and long enough cable to allow comfortable use. 


\begin{table*}[t]
    \centering
    \begin{tabular}{ccccccccccc}
     &
     \rot{\textsc{WaveGlove}-single} &
    \rot{\textsc{WaveGlove}-multi} &
    \rot{MHEALTH} &
    \rot{USC-HAD} &
    \rot{UTD-MHAD1} &
    \rot{UTD-MHAD2} &
    \rot{WHARF} &
    \rot{WISDM} &
    \rot{uWave} &
    \rot{Average}
    \\ \hline
    
    \multicolumn{1}{c|}{
        \begin{tabular}[c]{@{}c@{}}
            Baseline 
            Decision Tree
        \end{tabular}
    } & 
    \multicolumn{1}{c|}{99.10} &
    \multicolumn{1}{c|}{96.63} &
    \multicolumn{1}{c|}{\textbf{93.41}} &
    \multicolumn{1}{c|}{89.22} &
    \multicolumn{1}{c|}{67.51} &
    \multicolumn{1}{c|}{81.90} &
    \multicolumn{1}{c|}{66.26} &
    \multicolumn{1}{c|}{61.61} &
    \multicolumn{1}{c|}{70.72} &
    \multicolumn{1}{c|}{80.71} 
    \\ \hline 
    
    \multicolumn{1}{c|}{
        \begin{tabular}[c]{@{}c@{}}
            DeepConvLSTM 
            (2016)\cite{deepconvlstm}
        \end{tabular}
    } &
    \multicolumn{1}{c|}{98.05} &
    \multicolumn{1}{c|}{99.30} & 
    \multicolumn{1}{c|}{81.01} & 
    \multicolumn{1}{c|}{83.97} & 
    \multicolumn{1}{c|}{67.29} & 
    \multicolumn{1}{c|}{86.50} & 
    \multicolumn{1}{c|}{67.98} & 
    \multicolumn{1}{c|}{\textbf{91.23}} & 
    \multicolumn{1}{c|}{98.10} &
    \multicolumn{1}{c|}{85.94} 
    \\ \hline
           
    \multicolumn{1}{c|}{
        \begin{tabular}[c]{@{}c@{}}
            DCNN Ensemble 
            (2019)\cite{dcnn_ensemble}
        \end{tabular}
    } & 
    \multicolumn{1}{c|}{-} &
    \multicolumn{1}{c|}{-} &
    \multicolumn{1}{c|}{93.09} & 
    \multicolumn{1}{c|}{88.49} & 
    \multicolumn{1}{c|}{62.03} & 
    \multicolumn{1}{c|}{81.63} & 
    \multicolumn{1}{c|}{75.50} & 
    \multicolumn{1}{c|}{89.01} & 
    \multicolumn{1}{c|}{-} &
    \multicolumn{1}{c|}{-} 
    \\ \hline 

    \multicolumn{1}{c|}{
    \begin{tabular}[c]{@{}c@{}}
        Transformer-based 
        (proposed)
    \end{tabular}
    } &
    \multicolumn{1}{c|}{\textbf{99.40}} &
    \multicolumn{1}{c|}{\textbf{99.99}} &
    \multicolumn{1}{c|}{90.35} &
    \multicolumn{1}{c|}{\textbf{89.83}} &
    \multicolumn{1}{c|}{\textbf{76.32}} &
    \multicolumn{1}{c|}{\textbf{88.42}} &
    \multicolumn{1}{c|}{\textbf{78.63}} &
    \multicolumn{1}{c|}{84.53} &
    \multicolumn{1}{c|}{\textbf{98.80}} &
    \multicolumn{1}{c|}{\textbf{89.59}} 
    \\ \hline 
    \end{tabular}
\caption{
    Accuracy comparison of classification methods.
    The "-" symbol denotes that the given method was not applied on the dataset.
    Note that this table only showcases the most relevant methods and datasets. For the full set of results (over $12$ methods and $11$ datasets),
    exact model and hyperparameter description, as well as the code used to preprocess the input datasets and produce the results, please refer to
    \href{https://github.com/Zajozor/waveglove}{https://github.com/zajozor/waveglove}.
}
\label{tab:stats}
\end{table*}

\subsection{\textsc{WaveGlove} datasets}

\begin{figure}
    \centering
    \includegraphics[width=\linewidth]{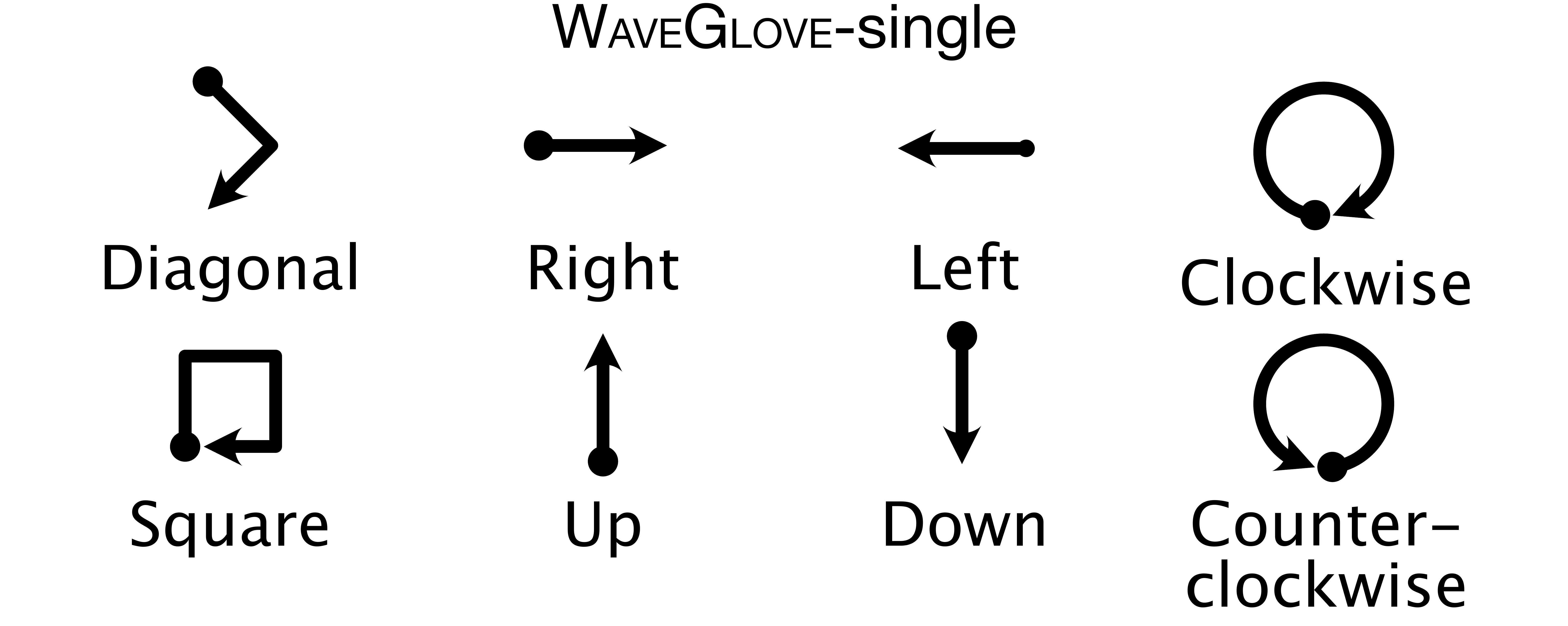}
    \caption{Eight gestures from the \textsc{WaveGlove}-single vocabulary.}
    \label{fig:single-vocab}
\end{figure}

\begin{figure}
    \centering
    \includegraphics[width=\linewidth]{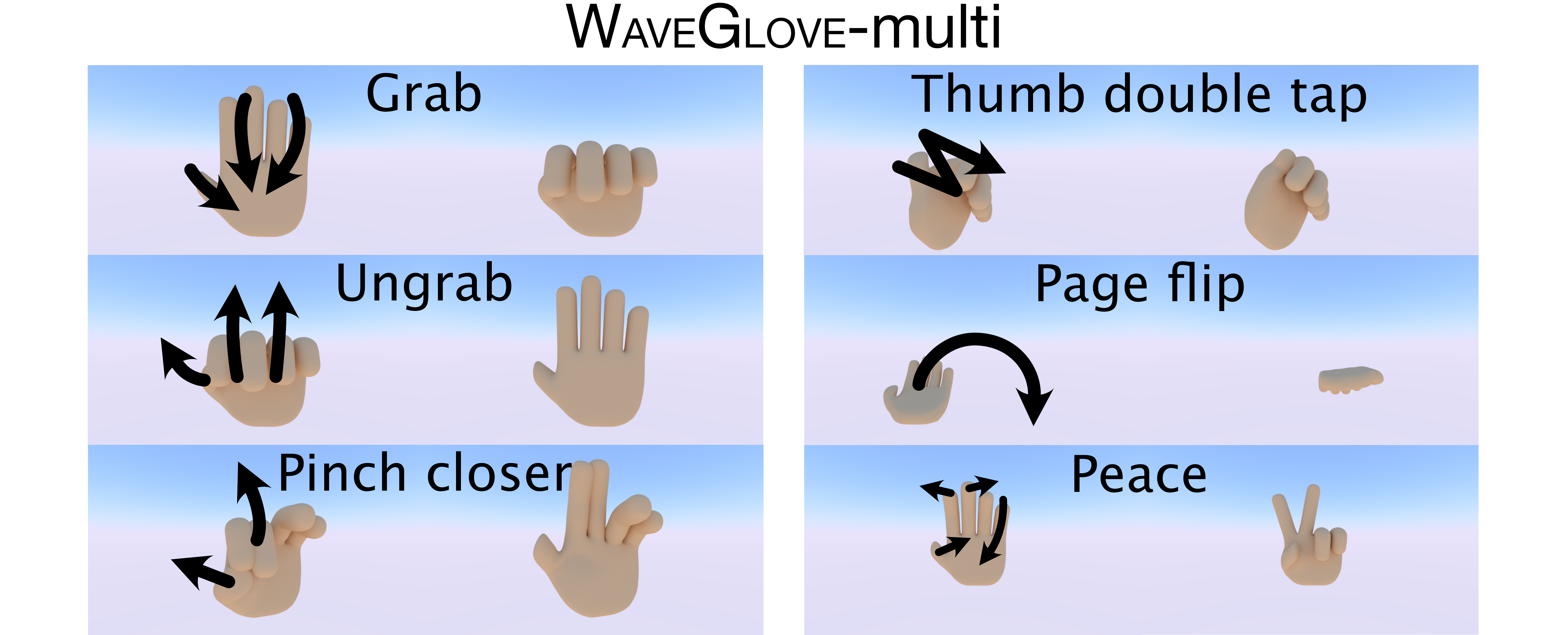}
    \caption{A selection of the gestures from the \textsc{WaveGlove}-multi vocabulary, which contains $10$ gestures in total.}
    \label{fig:multi-vocab}
\end{figure}

Two gesture vocabularies were used to acquire two separate datasets.

The \textsc{WaveGlove}-single vocabulary consists of $8$ gestures and a null class -- the state in which no gesture is performed. These gestures are simple movements, during which the whole hand performs a uniform motion (Figure \ref{fig:single-vocab}).
This set of gestures is motivated by \cite{gestures-identified},
which identified it to be a popular and user-preferred set for device control.
It is also one of the most often used gesture vocabularies
\cite{06_costante_dataset, 05_uwave, 18_egyptian}.
\textsc{WaveGlove}-single is primarily used as a reference point --
it uses the same set of gestures as previous work, with the only
difference being the use of five sensors instead of a single one.

On the other hand, \textsc{WaveGlove}-multi was designed as a set of custom hand gestures
which would showcase the potential of multiple sensors. These gestures consist of different movements
for individual fingers and some gestures differ only in specific finger movements.
This vocabulary is used to demonstrate that
the use of multiple sensors located on the hand significantly expands the space of
recognizable gestures and allows the hand gestures to become much more natural.
It contains $10$ gestures and a null class.

The key difference between the two gesture vocabularies lies in their level of complexity. While gestures from \textsc{WaveGlove}-single were successfully classified using a single sensor in previous work, the gestures in \textsc{WaveGlove}-multi are designed to fully utilize multiple sensors.
We take a closer look at this difference in an ablation study in Section \ref{sec:ablation}.
 
Several important factors affect the final quality of the dataset~\cite{55_datasets}.
In order to make our samples more realistic, ensure the recordings stay independent and improve the reproducibility of our experiments, we took several measures.
The gestures are recorded in short (approximately 2 minute) sessions.
Each session consists of recording a few samples for each of the gestures in one of the datasets.
In each session, the gestures are presented in a random order and the recording software requires a break (of a few seconds) between any two gestures.
This improves gesture independence and allows the data captured during the breaks to be used as the null class.

Before recording, the test subjects are first presented with uniform instructions.
The test subjects themselves mark the beginning and the end of a sample by pressing and
releasing a button on the keyboard. In case the subject made a mistake in the process
(eg. forgot to move the hand, or pressed the button too late), they are allowed to re-record a sample.

In total, the \textsc{WaveGlove}-single dataset contains over $1000$ samples
and the \textsc{WaveGlove}-multi dataset contains over $10000$ samples.
To the best of our knowledge \textsc{WaveGlove} is the first device utilizing multiple inertial sensors used to record publicly available datasets of this scale.

\section{Dataset normalization}
\label{sec:datasets}

In order to ensure fair comparison of the presentend HAR and HGR models, we prepare a set of $11$ datasets as a standardized benchmark.

We include the $6$ datasets preprocessed in \cite{41_standardization_sota}
using the Leave-One-Trial-Out (LOTO) folds, which ensure that all samples
from a single trial are either in the training or the test set.
The LOTO folds use semi-overlapping windows for sampling, while
the rest of the datasets are sampled using non-overlapping windows.
These $6$ datasets include MHEALTH and USC-HAD -- two commonly used HAR datasets.
The uWave~\cite{05_uwave} dataset is a single-sensor HGR dataset
with the same vocabulary as \textsc{WaveGlove}-single.
%
Finally, we include
the Opportunity~\cite{ds-opportunity1} and
Skoda~\cite{ds-skoda1} datasets,
along with the two \textsc{WaveGlove} datasets, resulting in $11$ datasets
in total.

\section{Methods}
\label{sec:results}

For classification we use methods ranging from Classical
Machine Learning to approaches based on Deep Learning.
To make our results comparable, 
we reproduce several previously published methods.
We further propose a classification method which utilizes the self-attention mechanism
and is based on the Transformer.
A Transformer-based architecture was utilized for HAR in~\cite{32_self_attention},
upon which we further improve by simplification and by using a more commonly used dot-product attention.

\begin{figure}[t]
  \centering
  \includegraphics[width=0.64\linewidth]{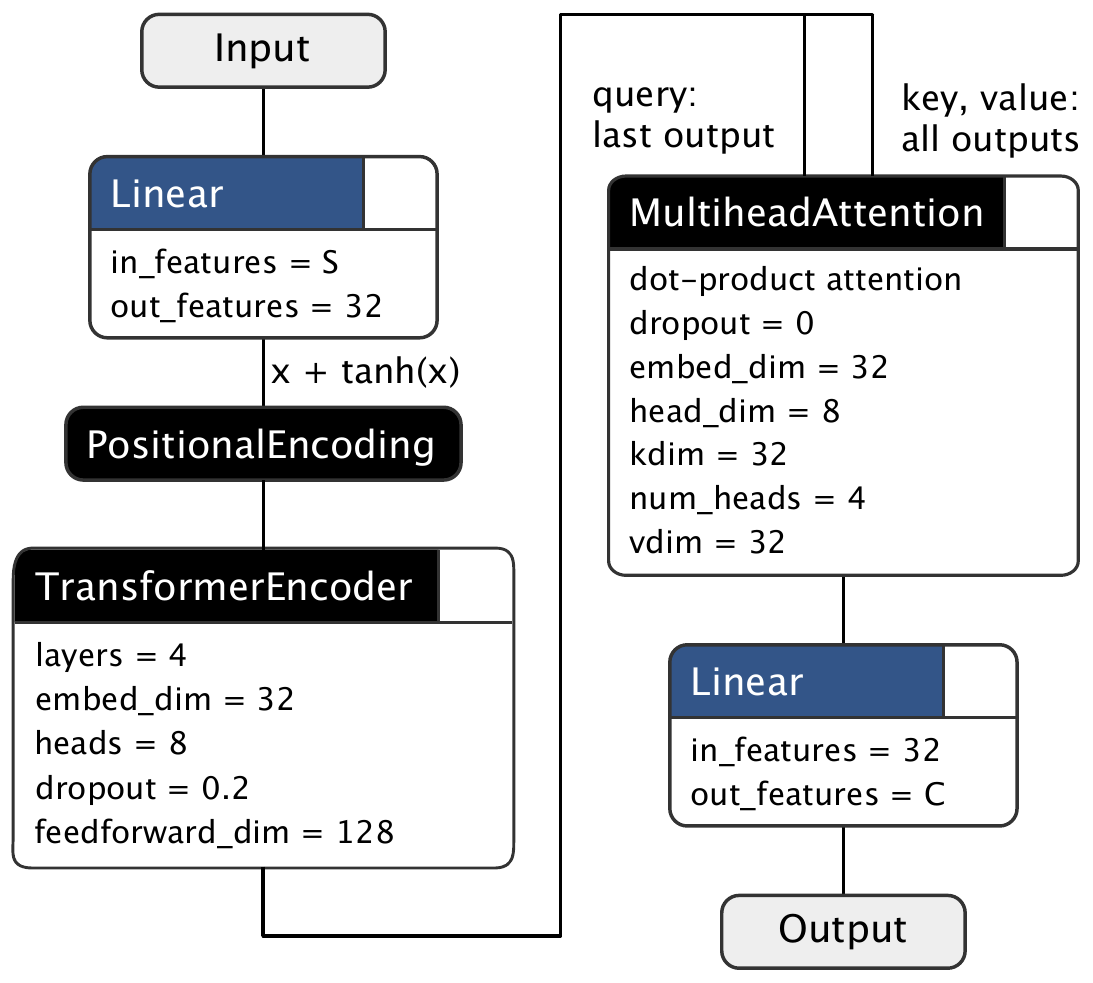}
  \caption{
  Architecture of the proposed Transformed-based model.
  }
  \label{fig:model-architecture}
\end{figure}

Figure~\ref{fig:model-architecture} shows the proposed architecture.
The first linear layer transforms an input of size $B \times T \times S$
($B=$ batch size, $T=$ temporal length, $S=$ sensor channels)
to $B \times T \times 32$, creating a constant-size sensor embedding.
Next, positional encoding\cite{attention-is-all} and $4$
Transformer encoder layers are used to provide an input ($T \times B \times 32$) for a dot-product
attention layer (the final encoder output in the temporal dimension is used as the query).
The final linear layer yields a one-hot encoding for target classes
($B \times 32 \rightarrow B \times C$, $C=$ classes). 

A hyperparameter sweep has been performed yielding the final, best-performing parameters.
Final model accuracy comparison can be found in Table~\ref{tab:stats}.

\begin{figure*}[t]
  \centering
  
  \subfigure{
  \includegraphics[width=0.45\linewidth]{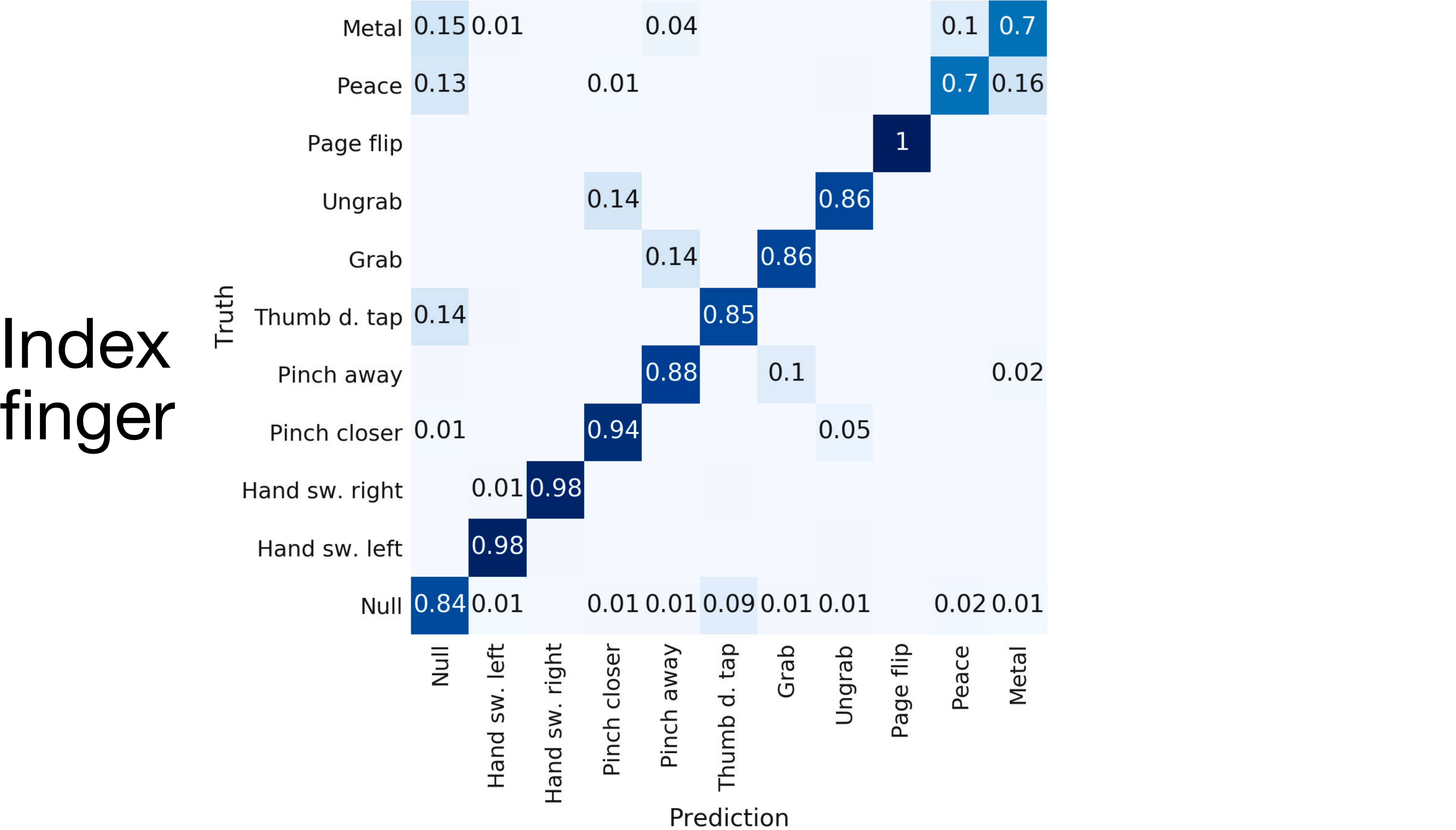}
  }\hfill
  \subfigure{ \includegraphics[width=0.45\linewidth]{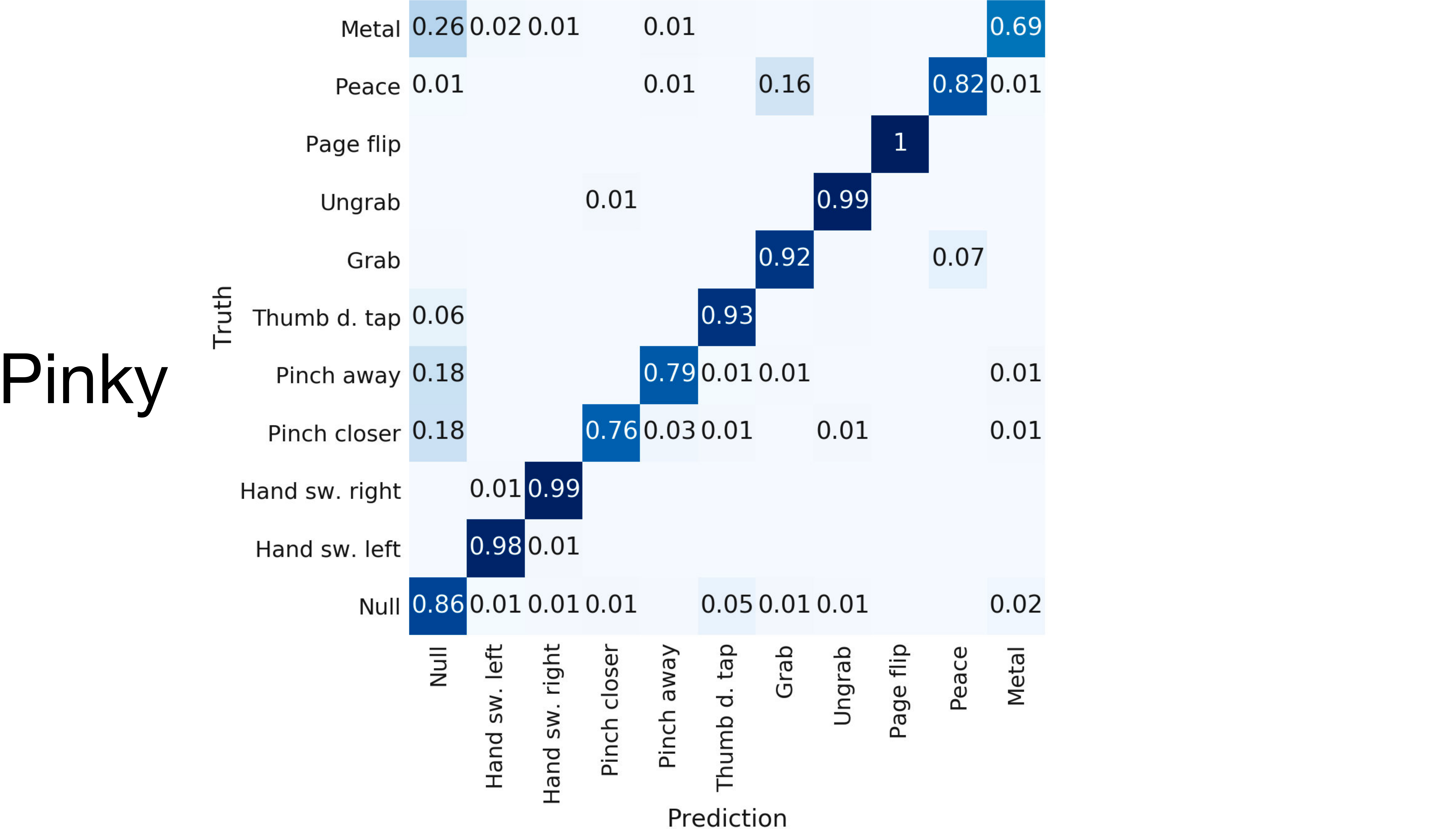}
  }
  \caption{Confusion matrices of a baseline model trained solely on \textbf{index finger} (left) and \textbf{pinky} (right) sensor data}
   \label{fig:index-pinky}
\end{figure*}

\section{Ablation study}
\label{sec:ablation}

Most of the evaluated methods achieved near-perfect
accuracy on \textsc{WaveGlove} datasets.
Given the multi-sensor setup, this opens
up an opportunity to conduct an ablation study to examine the
effect of various factors on the final performance.

\subsection{Finger attribution}

In Figure \ref{fig:index-pinky} we compare the confusion matrices
obtained by training a model on data from only a single sensor.
It allows us to observe a few distinguishable patterns.

Using the index finger sensor, the
\textit{Metal} and \textit{Peace} gestures are classified as the \textit{Null} class since they do not include index finger movement.
The gesture pairs \textit{Ungrab}/\textit{Pinch closer} and
\textit{Grab}/\textit{Pinch away} include the same index finger movement,
resulting in higher inter-class confusion.
When using the pinky sensor,
the \textit{Peace} and \textit{Grab} gestures are often confused.
Several gestures are confused with the \textit{Null} class, because they do not involve
significant pinky movement.

This experiment helps us demonstrate that sensor placement on different fingers
can have significant effect on classification results, particularly when
various fingers are featured prominently in the gestures.

\subsection{Impact of multiple sensors}

\begin{figure}[h]
  \centering
  \includegraphics[width=\linewidth]{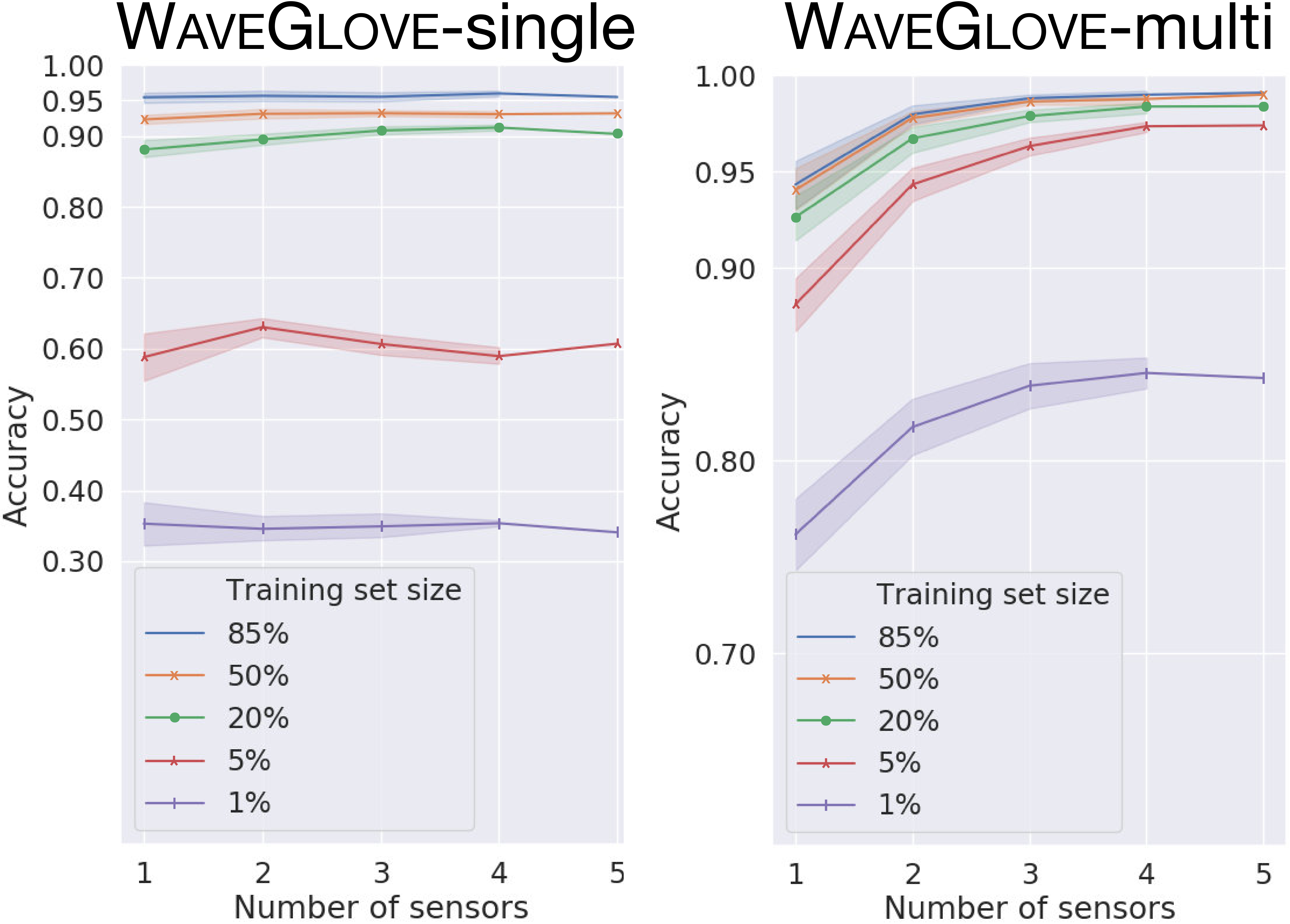}
  \caption{
  \hspace{0cm}
    Average accuracy based on the amount of sensors used and the training set size
    as measured on the two \textsc{WaveGlove} datasets.
  }
  \label{fig:attribution}
\end{figure}

Figure \ref{fig:attribution} depicts how the number of used sensors and the
training set size affect the final accuracy\footnote{When there are multiple combinations
for a certain amount of sensors, we plot the mean accuracy and the sleeves show the range.}.
With \textsc{WaveGlove}-single, the accuracy does not increase with the
addition of multiple sensors. This is what we would expect, as the
gestures in this vocabulary do not make use of multiple sensors.

Conversely, for the \textsc{WaveGlove}-multi dataset we observe a clear
positive impact of using more sensors on the accuracy. The improvements can be considered
significant when using up to three sensors. Beyond that the increase in performance is marginal.
This positive impact is even more significant when the training set size is limited.

\section{Conclusions}
\label{sec:conclusion}

In this work we present a custom multi-sensor hardware prototype,
which we used to acquire a dataset of over $11000$ gesture instances.
To the best of our knowledge this is the largest publicly available dataset of multi-sensor hand gestures.
For a fair comparison with prior work we use $11$ HAR datasets, implement several previously published methods and compare their performance.
We further propose a Transformer-based network architecture, which
shows promising classification results across multiple datasets.

In an accompanying ablation study, we identify that (only) relevantly designed gestures benefit from the use of multiple sensors.
By comparing classification performance for two different sensor locations we demonstrate that the recognition accuracy has strong dependence on the location of the sensor and the type of classified gestures.

\vfill\pagebreak



\bibliographystyle{IEEEbib}
\bibliography{refs}

\end{document}